\documentclass{article}
\usepackage{graphicx} % Required for inserting images
\usepackage{xcolor}
\usepackage{amsmath}
\usepackage{subcaption} %for multiple figures in a figure
\usepackage{multirow} %include multiple rows in a table 
\usepackage[most]{tcolorbox} %for textbox
\usepackage{setspace} %double space
\usepackage{multicol}
\usepackage{pdflscape}

\newtcolorbox{myquote}[1][]{%
    colback=black!5,
    colframe=black!5,
    notitle,
    sharp corners,
    borderline west={2pt}{0pt}{red!80!black},
    enhanced,
    breakable,
    }
\usepackage[citestyle=authoryear,natbib=true,backend=bibtex]{biblatex}
\bibliography{ref.bib}

\title{AI Revolution on Chat Bot: Evidence from a Randomized Controlled Experiment  }
\author{Sida Peng\thanks{Office of Chief Economist, Microsoft} \and Wojciech Swiatek\thanks{Microsoft} \and Allen Gao\thanks{Microsoft} \and Paul Cullivan\thanks{Microsoft}  \and Haoge Chang\thanks{Microsoft Research}}

\date{November 2023}

\begin{document}

\maketitle

\section{Introduction}
In recent years, generative AI has undergone major advancements, demonstrating  significant promise in augmenting human productivity. Notably, large language models (LLM), with ChatGPT-4 as an example, have drawn considerable attention. Many companies are now incorporating LLM-based tools within their organizations and integrating it with various products \citep{ChatGPTOrg,Forbes}. There is increasing interest on evaluating the impact of LLMs on human decision-making and productivity.

Numerous articles have examined the impact of LLM-based tools in lab settings or designed tasks \citep{peng2023impact, spatharioti2023comparing,noy2023experimental,dell2023navigating}, or in observational studies \citep{brynjolfsson2023generative}. In these investigations, LLM-based tools are deployed to aid humans in various tasks, with measured outcomes including task completion time and accuracy. It is generally observed that LLM-based tools are able to increase users' productivity substantially.

Despite recent advances, field experiments applying LLM-based tools in realistic settings are limited. This paper presents the findings of a field randomized controlled trial assessing the effectiveness of LLM-based tools in providing unmonitored support services for information retrieval. While superior service quality is expected from LLM-based tools, concerns such as hallucinations raise questions about their effectiveness. Consequently, an empirical investigation is necessary.

\begin{figure}[h]
    \centering
  \includegraphics[width=1\linewidth]{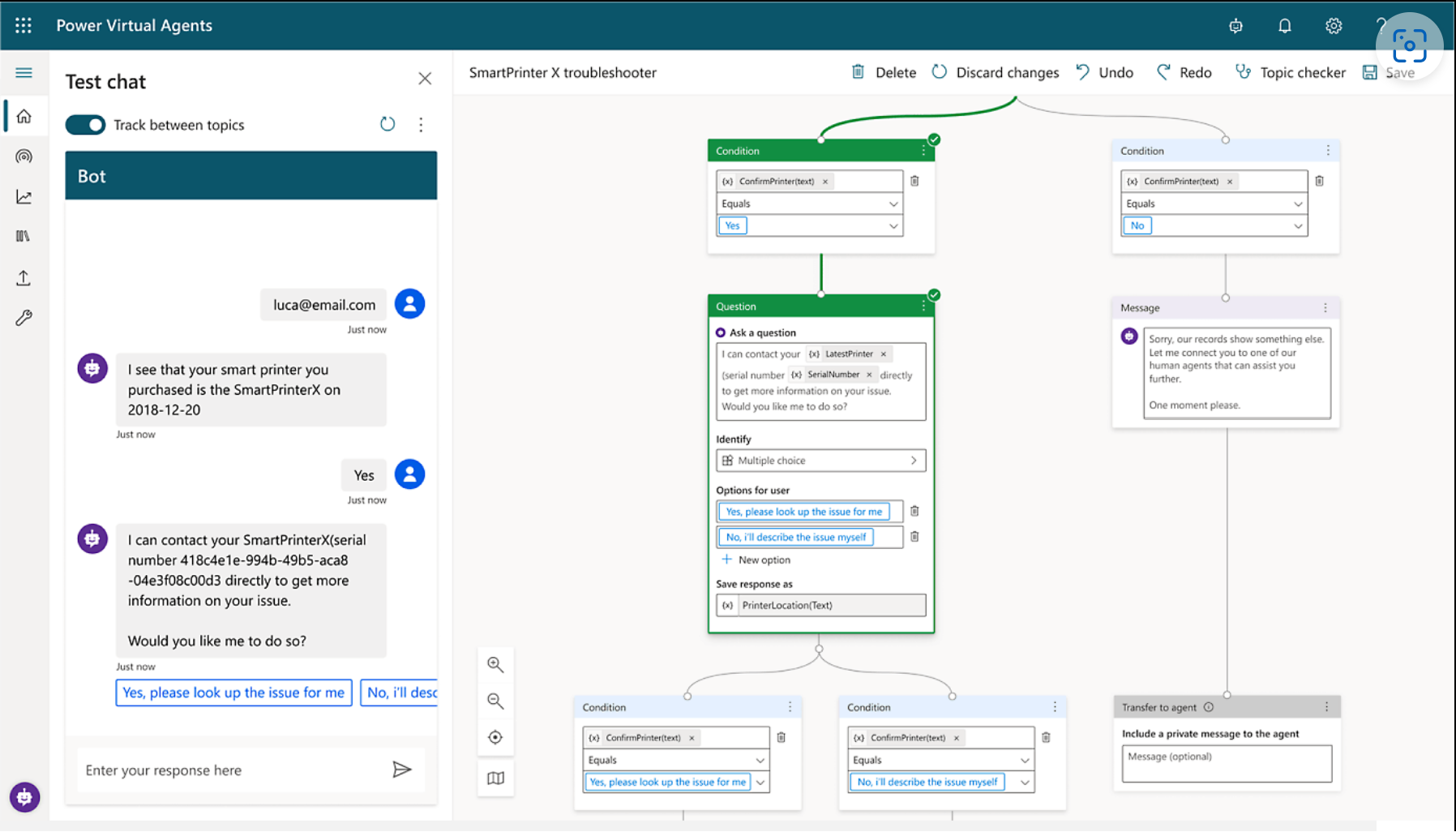}
    \caption{Power Virtual Agents: Flow Chart Based System}
    \label{fig:powerVA}
\end{figure}

We collaborated with a team managing support chat bots that support Microsoft's internal developers. Prior to adopting GPT-based models, these bots operated on a flowchart-based system called Power Virtual Agents (PowerVA, see Figure \ref{fig:powerVA}). Users navigated through predefined categories to find document links potentially relevant to their queries.

In our experiment, we integrated a particular support bot, the Work Management Support Bot, with GPT-based tools and compared its performance to the existing keyword-based flowchart support bot. This bot is tailored to assist Microsoft software developers with login and access issues. The new GPT-based approach features a bot (hereafter GPT-based bot) that enables users to ask questions in natural language and receive direct answers from the same document sources used by the flowchart support bot (hereafter classical bot).

The primary outcome of interest in our study is the escalation decision, which is defined by a user’s decision to escalate the inquiry and seek support from the back-end engineer. A good support bot should lead to low escalation rates. Our experiment suggests that the GPT-based bot reduces the escalation rate by 9.2 percentage compared with the classical bot. This represents a 53.8 percent reduction in escalation rate relative to the baseline of the classical bot.

In addition, we implemented two versions of the GPT-based support bots, one based on the GPT4 model and the other based on the GPT3.5 model. We compare their escalation rates and token consumption. While we find no significant difference in escalation rates between the two models, there are notable differences in token usage: although GPT4-based bot consumes less tokens per question on average, GPT3.5-based bot has a price advantage under the current pricing structure of GPT models \citep{Pricing}.   

Our results add to the literature of experimental and observational research on the productivity-enhancing effects of LLM-based tools. For instance, \cite{noy2023experimental} examines the individual and distributional productivity effects of ChatGPT on writing task completion and quality among experienced, college-educated professionals.  \cite{peng2023impact} investigates how GitHub Copilot affects software developers' productivity in programming tasks.  \cite{dell2023navigating} randomizes GPT4 tools to consultants and measures the change in productivity. \cite{spatharioti2023comparing} evaluates LLM-based tools in aiding consumers with search tasks. Finally, \cite{brynjolfsson2023generative} studies the impact of AI tools on productivity in an observational setting using a difference-in-differences approach. 

The rest of the paper is as follows. Section 2 introduces backgrounds and experimental designs. Section 3 contains results of the experiments, and Section 4 includes details on the specifications of our statistical analysis.

\section{Backgrounds and Experimental Designs}

\begin{landscape}
\begin{figure}[h]
\centering
\begin{subfigure}{.5\textwidth}
  \centering
  \includegraphics[width=0.8\linewidth]{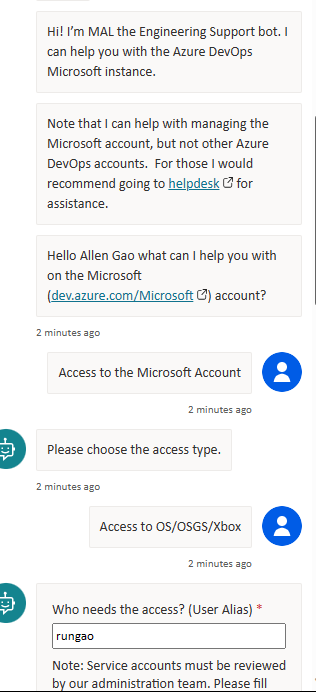}
  \label{fig:old1}
\end{subfigure}%
\begin{subfigure}{.5\textwidth}
  \centering
  \includegraphics[width=0.8\linewidth]{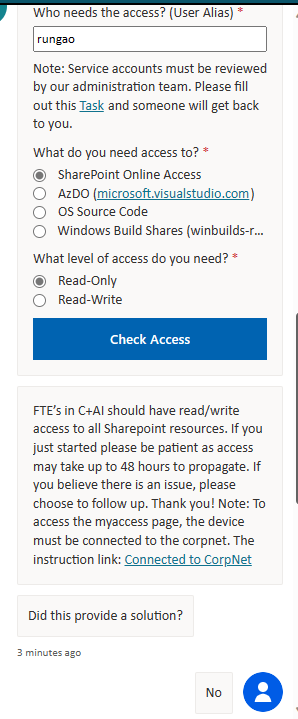}
  \label{fig:old2}
\end{subfigure}%
\begin{subfigure}{.5\textwidth}
  \centering
  \includegraphics[width=0.8\linewidth]{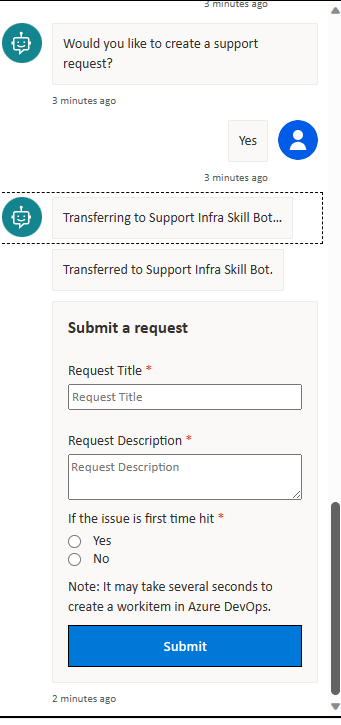}
  \label{fig:old2}
\end{subfigure}
\caption{Microsoft PowerVA-based Support Bot Experience}
\label{fig:old}
\end{figure}

\end{landscape}

\begin{landscape}
\begin{figure}[h]
\centering
\begin{subfigure}{.5\textwidth}
  \centering
  \includegraphics[width=0.8\linewidth]{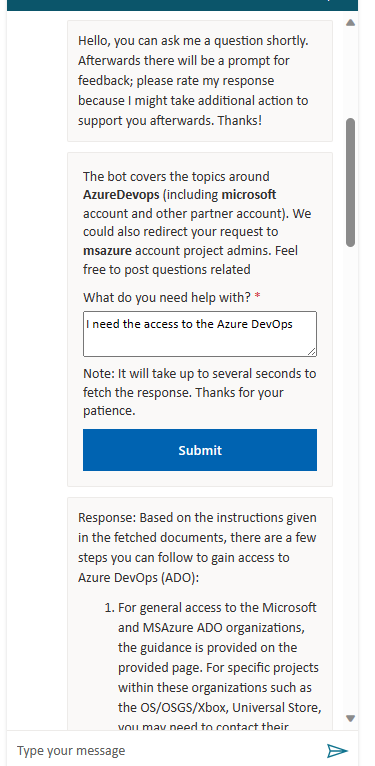}
  \label{fig:new1}
\end{subfigure}%
\begin{subfigure}{.5\textwidth}
  \centering
  \includegraphics[width=0.8\linewidth]{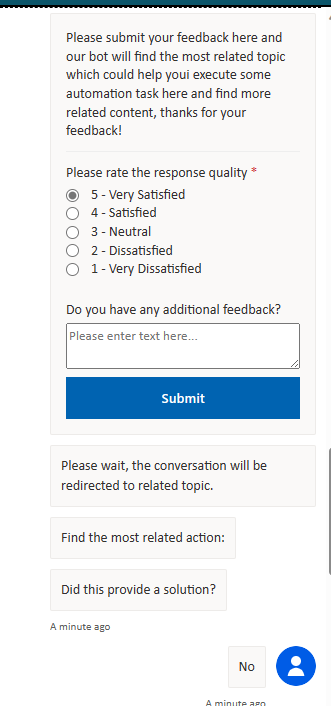}
  \label{fig:new2}
\end{subfigure}%
\begin{subfigure}{.5\textwidth}
  \centering
  \includegraphics[width=0.8\linewidth]{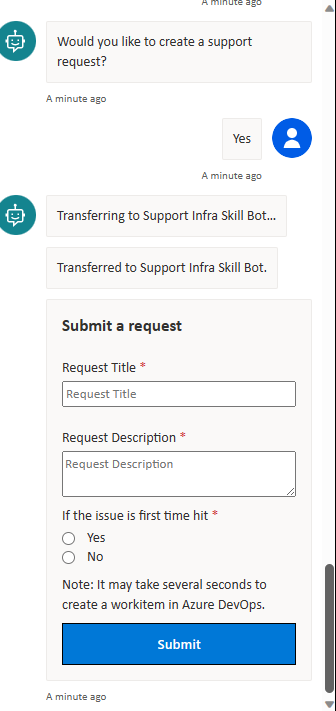}
  \label{fig:new3}
\end{subfigure}
\caption{GPT-based Support Bot Experience}
\label{fig:new}
\end{figure}
    
\end{landscape}

We conducted a randomized control trial (RCT) to compare the performance of the Power Virtual Agents (classical) based bot with the GPT-based bot. This chatbot assists Microsoft developers with login and access issues. All traffic to the chatbot service was randomized into control and treatment groups. The control group interacted with the existing bot, based on the PowerVA service, offering a flowchart-style experience where users navigate predefined options to find documents addressing their questions. The treatment group used the new GPT-based bot, which allows users to pose questions in natural language and provides direct answers from the same document sources as the classical bot. Figure \ref{fig:old} depicts the PowerVA-based support bot experience, and Figure \ref{fig:new} illustrates the GPT-based support bot experience.

If developers are unsatisfied with the bot's responses, they have the option to escalate their questions for real human intervention. Our back-end engineers assign 10 minutes to resolve each escalated case, and most cases are resolved within this estimate. The traffic varies between 5 to 20 cases per week, depending on seasonality. One may witness heavier-than-usual traffic during Monday, specific months, and after reorganizations when some users lose access and turn to the bot to request new permissions.

Our experiments had two waves. The first wave took place between May 05, 2023 and July 21, 2023, and the second wave from July 21, 2023 to Oct 12, 2023. In the first wave, users in each session were randomly assigned to either the classical bot or the GPT4-based version. In the second wave, a new GPT3-supported version was introduced, and and users were randomly assigned to one of three options: the classical bot, the GPT4-based bot, or the GPT3.5-based bot. No additional instructions were provided on how users should interact with the support bots.

For each session, we collected data such as session ID, starting time of the session, engagement decision, duration of the engagement, and escalation decision. For the GPT-based versions, we were able to collect data such as users' first prompts and support bot's first responses.

The primary outcome of our study is the escalation decision, which reflects a user's choice to escalate their inquiry and seek assistance from a back-end engineer. An effective support bot should accurately comprehend a user's question, retrieve the correct information, and provide clear responses. Ultimately, a high-quality support bot should result in a low escalation rate, thereby reducing the workload for back-end engineers in supporting users.

\section{Results}
We collected data on 3296 sessions over a span of five months. There are 1413 engaged cases and 165 escalations. We calculate the overall engagement rate as:
\begin{equation*}
    \text{Overall Engagement Rate} = \frac{1413}{3296} = 42.9\%.
\end{equation*}
Among the engaged cases, the overall escalation rate is calculated as
\begin{equation*}
     \text{Overall Escalation Rate} = \frac{165}{1413} = 11.7\%.   
\end{equation*}
\subsection{Engagement Rates}\label{section:engagementrates}

Users may not to engage with the support bot services for various reasons. For instance, the support bots automatically initiate a login action when the conversation starts. This login process can take anywhere from 5 to 30 seconds, during which users might refresh the page, inadvertently skipping the existing session. Additionally, some users may accidentally click on the wrong page, leading to unintended visits to the support bot service.

In our analysis, we concentrate on the outcomes of engaged sessions to evaluate the quality of the support bots. A GPT-based bot session is deemed engaged if the user posts a question and the bot responds. For a classical bot session, engagement is defined as the user utilizing at least one functionality of the bot. Our findings indicate no significant difference in engagement rates between classical sessions and GPT-based sessions (difference-in-means = -0.026, t-statistics = -1.457, p-value = 0.145).

\subsection{Primary Outcome: Escalation Rate}\label{section:escalation}

We observe 66 escalations out of 835 engaged sessions with the GPT-based support bots, and 99 escalations out of the 578 engaged sessions with the classical support bots. 

We calculate the average escalation rate as: 
\begin{equation*}
    \text{Average Escalation Rate (GPT-based bot)} = \frac{66}{835} = 7.9\%
\end{equation*}
\begin{equation*}
    \text{Average Escalation Rate (classical bot)} = \frac{99}{578} = 17.1\%
\end{equation*}
This is a significant 9.2 percentage point reduction (t-statistics=-5.05, p-value=4.9e-07) in the average escalation rate when using the GPT-based support bots. 

In a relative term, this represents a 53.8 percent (9.2/17.1) reduction in average escalation rate compared to the baseline escalation rate of 17.1 percentage points for the classical support bot.

\subsection{Comparing  GPT3.5 and GPT4}\label{section:GPT3andGPT4}
In the second wave of our experiment and after July 21st 2023, we increased the number of sessions that are assigned to the GPT-based support bot. Further, we randomized sessions into either GPT3.5-based support bot or GPT4-based support bot. During the period between July 21st, 2023 and October 11th, 2023, we collected information on 190 engaged cases for the GPT4-based support bot and 203 engaged cases for the GPT3.5-based support bot. 

Among the 203 engaged cases for the GPT3.5-based version, there were 17 escalations, and among the 190 engaged cases for GPT4-based version, there were 19 escalations. The average escalation rates for the GPT3.5-based and GPT4-based versions are
\begin{equation*}
    \text{Average Escalation Rate (GPT3.5-based bot)} = \frac{17}{203} = 8.4\%
\end{equation*}
\begin{equation*}
    \text{Average Escalation Rate (GPT4-based bot)} = \frac{19}{190} = 10\%
\end{equation*}

The escalation rate of the GPT4-based version is slightly higher than that of the GPT3.5-based version, but the difference is not statistical significant (t-statistics=0.556, p-value=0.579). 

We also compared the GPT3.5-based and GPT4-based versions in terms of input and output token consumption. Input and output tokens form the basis for cost calculations when using GPT-based services \citep{Pricing}. Since the pricing plans for tokens vary depending on the version used, our comparison here is relevant to cost considerations.

\begin{table}[h]
    \centering
    \begin{tabular}{|c|c|c|}
\hline
    & Input Tokens & Output Tokens \\
\hline
  GPT3.5-based  bot     & 2575.0& 190.3 \\
\hline
  GPT4-based bot       & 2018.4 & 235.5\\
\hline
    \end{tabular}
    \caption{Average token consumptions for GPT3.5-based and GPT4-based bots.}
\label{table:tokenassumption}
\end{table}

\begin{figure}
    \centering
\begin{subfigure}{.5\textwidth}
    \centering
    \includegraphics[width=1.3\linewidth]{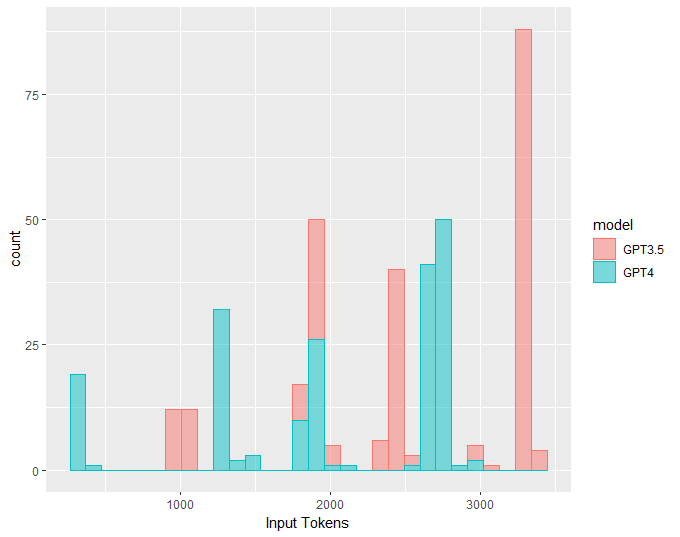}
    \label{fig:inputtokens}
\end{subfigure}\\
\begin{subfigure}{.5\textwidth}
    \centering
    \includegraphics[width=1.3\linewidth]{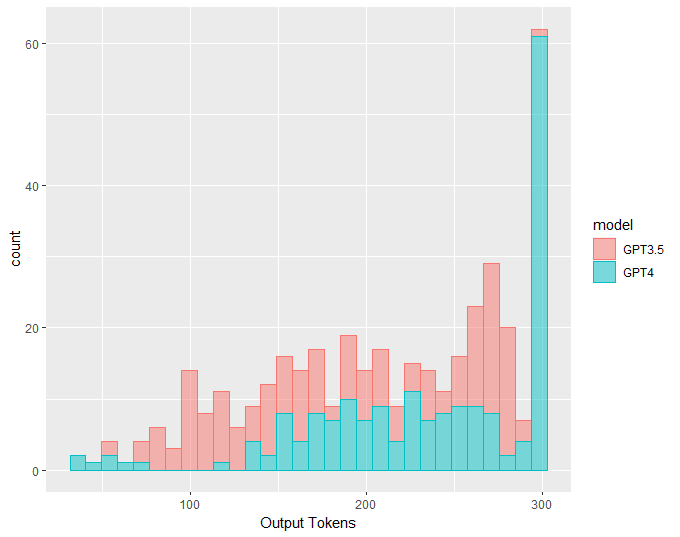}
    \label{fig:outputtokens}
\end{subfigure}\\
\begin{subfigure}{.5\textwidth}
    \centering
    \includegraphics[width=1.3\linewidth]{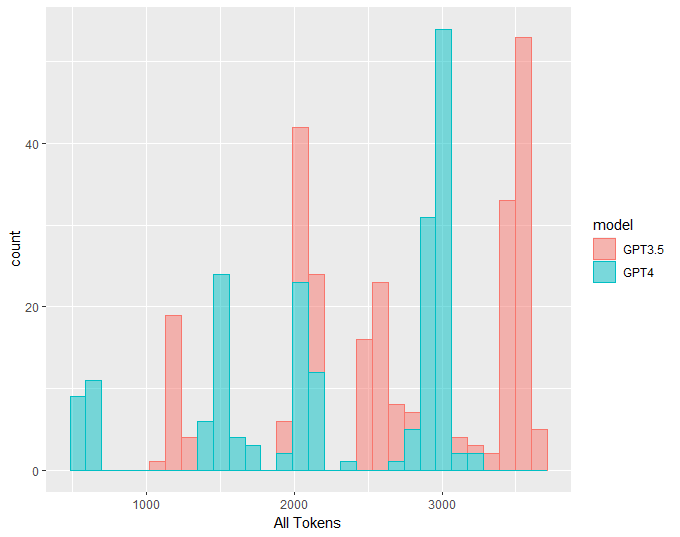}
    \label{fig:outputtokens}
\end{subfigure}\\
\caption{Distributions of Token Consumption for GPT3.5-based and GPT-4 based bots}
\label{fig:token}
\end{figure}

We found that users' questions consume from 290 to 3363 input tokens, from 38 to 300 output tokens, and from 500 to 3620 total tokens (input tokens + output tokens). Figure \ref{fig:token} plots token consumption for both GPT3.5-based and GPT4-based bots. As shown in Table \ref{table:tokenassumption}, on average the GPT3.5-based bot consumes more input tokens than the GPT4-based bot does (difference-in-means=-556.6, t-value=-6.864, p-value=2.6e-11). The GPT4-based bot consumes more output tokens than the GPT3.5-based bot (difference-in-means=45.2, t-value=0.557, p-value=0.578), but this difference is not statistically significant.

The variation in input token consumption between the two versions can be explained by the fact that GPT4 is able to process several frequently-referenceed documents with less resources. The difference of the output tokens can be explained by the fact that GPT4-based version often outputs longer but more structured responses. For example, for the same prompt "I need access", GPT4-based bot outputs

\begin{myquote}
To request access, you need to follow the appropriate steps based on the type of access you require. Here's a general outline of the process:
\begin{enumerate}
    \item Determine the type of access you need:
\begin{itemize}
    \item Access Type 1
    \item Access Type 2
    \item ...
\end{itemize}
    \item Visit the [NAME](URL) portal.
    \item Depending on the type of access you need, follow the appropriate link(s) provided in the documentation to request access.
    \item Fill out the required information and submit your request.
    \item Wait for approval from the appropriate authority based on your organization and the type of access you're requesting.
    \item Please refer to the documentation for specific links and approval requirements based on your organization and the type of access you need. If you have any issues or need further assistance, contact the [NAME](URL). 
\end{enumerate} 
\end{myquote}
For the GPT3.5-based bot, it outputs
\begin{myquote}
Hi there! You need access to resources that are managed by Microsoft. To get access, you need to go through [NAME](URL). [ServiceName] is supported by [Name](URL). 

For work item/resource access for [Projects] that are not [Projects], you can ask [NAME](URL) to determine who that is. 

For full legacy source access, you can apply [NAME](URL). For legacy test sources only, you can apply [NAME](URL). For legacy branch owners' publics, you can apply [NAME](HERE)

\end{myquote}

We can calculate average cost per question for both GPT-3.5 based bot and GPT4-based bot, using the existing pricing schedule \citep{Pricing} (price based on Azure public price at Dec 2023). Based on the information in Table \ref{table:tokenassumption}, the average cost per question is 7.5 cents (0.03*2018.4/1000 + 0.06*235.5/1000) for GPT4-based bot and 0.3 cents (0.0010*2575/1000 + 0.0020*190.3/1000 ) for GPT3.5-based bot. Provided that the GPT3.5-based bot and GPT4-bot are able to provide similar experience, GPT3.5-based bot offers a more cost-efficient alternative compared with the GPT4-based bot.

\subsection{Robustness Check}\label{section:robustness}

Our randomization happens on a per-session level, so a user who use multiple sessions may see both the GPT-based and classical versions with the same question. We'll use a different experimental design to prevent this complication in the next round. In this session, we conduct a robustness check, specifically looking at users' escalation decisions in the first session.

We recorded 679 sessions with user alias between September 12, 2023 to October 12, 2023. There are 310 engaged cases during this time period and 263 cases are first sessions of users' interactions with the bots. There are 160 engaged GPT-based bot sessions with 10 escalations, and 103 Classical bot sessions with 21 escalations.  We calculate the average escalation rate as: 
\begin{equation*}
    \text{Average Escalation Rate (GPT-based bot)} = \frac{10}{160} = 6.3\%
\end{equation*}
\begin{equation*}
    \text{Average Escalation Rate (classical bot)} = \frac{21}{103} = 20.4\%
\end{equation*}
This is a significant 14.1 percentage point reduction (t-statistics=-3.19, p-value=0.001) in the average escalation rate when using the GPT-based support bots. It appears that results based on the restricted samples are qualitatively similar to the results reported in Section \ref{section:escalation}.

\section{Regression Details}

We used the statistical programming language R to analyze the data. The standard errors of all regression results are calculated using the HC2 formula, implemented in the \textit{sandwich} \citep{zeileis2019package} package in R. The t-tests are calculated using the \textit{lmtest}  \citep{hothorn2015package} package in R .

\subsection{Engagement Rates}\label{section:engagementrates_detail}
The reported results in Section \ref{section:engagementrates} are based on the regression specification:
\begin{equation*}
    \text{Engagement} = \beta_0 + \beta_1 \text{Version} + \epsilon, 
\end{equation*}
where \textit{Engagement} is a binary variable, with 1 indicating that the user engaged with the support bot and 0 otherwise. \textit{Version} is also a binary variable, with 1 representing a GPT-based support bot and 0 representing the classical version of the support bot.
\subsection{Escalation Rates }
The reported results in \ref{section:escalation} are based on the regression specification:
\begin{equation*}
    \text{Escalation} = \beta_0 + \beta_1 \text{Version} + \epsilon, 
\end{equation*}
where \textit{Escalation} is a binary variable, with 1 indicating the user escalated the issue to a back-end engineer and 0 otherwise. \textit{Version} is defined as in Section \ref{section:engagementrates_detail}. We limited our analysis to engaged sessions.
\subsection{Comparing GPT3.5 and GPT4}
The reported results in \ref{section:escalation} are based on the regression specification:
\begin{equation*}
    \text{Escalation} = \beta_0 + \beta_1 \text{GPT4} + \epsilon, 
\end{equation*}
where \textit{Escalation} is a binary variable, with 1 indicating that the user has escalated the issue to a back-end engineer and 0 otherwise. \textit{GPT4} is also a binary variable, with 1 representing a GPT4-based support bot and 0 indicating a GPT3.5-based bot.  We restricted samples to engaged sessions. The regression specifications for the input tokens and output tokens are similar but with different dependent variables.
\subsection{Robustness Check}
The reported results in Section \ref{section:robustness} use the same specifications as those reported in Section \ref{section:escalation}, with the constructed sample described in Section \ref{section:robustness}.

\printbibliography

\end{document}